\pdfoutput=1
\documentclass[preprint,preprintnumbers,amsmath,amssymb,superscriptaddress]{revtex4}
 
\newcommand{\be}{\begin{equation}}
\newcommand{\ee}{\end{equation}}
\newcommand{\ba}{\begin{eqnarray}}
\newcommand{\ea}{\end{eqnarray}}
\newcommand{\bd}{\begin{displaymath}}
\newcommand{\ed}{\end{displaymath}}

\newcommand{\commentout}[1]{{}}

\usepackage{amssymb}
\usepackage{color}
\usepackage{slashed}
\usepackage{amsmath}
\usepackage{graphicx}
\usepackage{epstopdf}
\usepackage{dcolumn}
\usepackage{bm}
\usepackage{hyperref}

\begin{document}

\title{MARTINI for heavy quarks: initialization, parton evolution, and hadronization with event generation}
\author{Clint Young}
\email{clinty@physics.mcgill.ca}
\affiliation{Department of Physics, McGill University, 3600 University Street, Montreal, Quebec, H3A\,2T8, Canada}
\author{Bj\"orn Schenke}
\email{bschenke@quark.phy.bnl.gov}
\affiliation{Physics Department, Bldg. 510A, Brookhaven National Laboratory, Upton, NY 11973, USA}
\author{Sangyong Jeon}
\email{jeon@physics.mcgill.ca}
\affiliation{Department of Physics, McGill University, 3600 University Street, Montreal, Quebec, H3A\,2T8, Canada}
\author{Charles Gale}
\email{gale@physics.mcgill.ca}
\affiliation{Department of Physics, McGill University, 3600 University Street, Montreal, Quebec, H3A\,2T8, Canada}

\date{\today}

\begin{abstract}

We present additions to the \textsc{martini} event generator for examining heavy quarks and quarkonia in heavy-ion collisions. 
All stages of a heavy-ion collision affect the observables associated with heavy quarks: the initial phase space of the heavy quarks are sampled 
with \textsc{pythia8.1}, the heavy quarks are evolved using Langevin dynamics and a 3+1-dimensional hydrodynamical description of the heavy-ion collision, and are fragmented and hadronized using a modified version of the color evaporation model that takes into account non-trivial evolution in position space, as well as the possibility of recombinant quarkonium production in heavy-ion collisions. We use this to re-examine the production of quarkonium at RHIC, and anticipating vertex detection we predict yields of $B_c$ mesons at RHIC and the LHC.

\end{abstract}

\maketitle

\section{Introduction}

Heavy quarks and quarkonia are hard probes of heavy-ion collisions; they are produced in hard processes in the initial moments of the collision and exist at all stages of the event. As a result, any calculation related to heavy quarks must consider nearly every aspect of heavy-ion physics: shadowing and anti-shadowing of the parton distribution functions, the perturbative cross sections, the finite-temperature evolution of heavy flavor, and finally, in-medium hadronization. Any calculation which neglects any of these steps and compares with an observable must quantitatively justify why the step was ignored.

As a result, the heavy quark and quarkonium observables in heavy-ion collisions now have a history beginning with qualitative understanding followed by scrutiny with simulation. As an example, consider the changes in $J/\psi$ yields caused 
by finite-temperature effects. These changes were first described as being caused by the non-Abelian analog of the photo-electric effect \cite{Shuryak:1978ij, Shuryak:1980tp}, and were estimated to be small. A large $J/\psi$ suppression was also predicted because of Debye screening of the $Q\bar{Q}$ potential, which had been seen in lattice calculations \cite{Matsui:1986dk}; the result suggested so large an effect that $J/\psi$ suppression was hoped to be the ``smoking gun" for formation of quark-gluon plasma in heavy-ion collisions. An ``anomalous" suppression was indeed observed by the collaborations at the SPS \cite{Abreu:1997jh, Alessandro:2004ap}. However, the results from RHIC indicate a more complicated suppression pattern than what can be described with the original results for large suppression \cite{Adler:2005ph}: the suppression at RHIC was small in comparison to expectations based on the results from SPS. Since then, heavy quark observables have been examined more carefully than had been done in the first semi-analytic predictions: the flow of heavy flavor has been estimated using hard thermal loop effective theory and a hydrodynamical description of heavy-ion collisions \cite{Moore:2004tg}; the rates for destruction and regeneration of 
$J/\psi$ particles at finite-temperature have been calculated using finite-temperature heavy quark effective theory and potentials from lattice QCD, and these rates were applied to the results at 
RHIC and the LHC \cite{Grandchamp:2002iy, Zhao:2008pp, Zhao:2011cv}; and finally, $J/\psi$ suppression was described with heavy quark diffusion and this description was integrated with a 2+1-dimensional hydrodynamical simulation for estimating changes in $J/\psi$ yields at RHIC 
\cite{Young:2008he, Young:2009tj}. Most recently, the rates for thermal quarkonium suppression were calculated away from thermal
equilibrium, for viscous plasmas, and applied to the suppression of bottomonium states at the LHC \cite{Strickland:2011mw}.

Event generation is a promising approach to keeping all of the steps under control while adding new physical considerations. In this paper, we 
describe modifications made to the \textsc{martini} event generator \cite{Schenke:2009gb}, which has been used successfully for examining 
jets and photons at RHIC and the LHC \cite{Young:2011qx, Dion:2011vd, Dion:2011pp}. We cover the physics of heavy quarks ``chronologically": in Section \ref{production} production, 
then in-medium evolution in Section \ref{evolution}, and finally hadronization in Section \ref{hadronization}. The new physical considerations of 
this paper are mostly in the section on hadronization, and their implementation in \textsc{martini} represents an important part of what Monte 
Carlo can do that is difficult analytically without many simplifying assumptions. We compute the results for observables at RHIC and the LHC in Section \ref{observables} and make a prediction of the yields of $B_c$ mesons, which are produced in heavy-ion collisions almost entirely recombinantly. Finally in the conclusions, areas where future work is needed are pointed out.

\section{The steps of heavy quark event generation}

\subsection{Initial production of heavy quarks}
\label{production}

Because of the masses of charm and bottom quarks, which are large compared with $\Lambda_{{\rm QCD} }$ and with the temperatures accessible in heavy-ion collisions, the processes producing these quarks have a large energy scale and can be described perturbatively. Heavy quarks are like high-energy jets in that their initial production in a heavy-ion collision is some elaboration of the production in proton-proton collisions. Also, the thermal 
production of charm and bottom quarks in heavy-ion collisions is likely small: t-channel gluon-gluon scattering which contributes most to the 
thermal heavy quark production is suppressed by two factors of the initial Bose-Einstein distributions of these gluons. For this reason, we 
consider only heavy quarks produced initially in the nuclear collision.

However, even perturbative production of heavy quarks is complicated by nuclear effects in heavy-ion collisions. First of all, the parton 
distribution functions in a heavy-ion collision are modified because of isospin. Every nucleon's parton distribution can be described with the 
weighted average of the neutron and proton distribution functions, $f^A_Z(x, Q) = \frac{Z}{A}f^p(x,Q)+\frac{A-Z}{A}f^n(x,Q)$, to take this effect into account. An effect with greater  significance for heavy quark production is the shadowing of the parton distribution functions by nucleons at the same point in the transverse 
plane, and the effect of anti-shadowing required by momentum conservation. These modifications have a satisfying description in color glass condensate effective theory, and if the saturation scale $Q_s > 2m_{H}$ the color glass condensate would be the best tool for describing heavy quark production at 
RHIC and the LHC \cite{Dominguez:2011cy}. Gluon saturation likely does cause some of the anomalous nuclear modification of $J/\psi$ production, however it cannot explain all of the observed modification. Another approach is more phenomenological and practical for event generation: parametrizing  the parton distribution functions for nucleons in bound states, and simply evolving these according to leading-twist DGLAP evolution \cite{Eskola:1998df}.

Another complication in the initial production arises at high-$p_T$ relative to the heavy quark's mass. In the factorized approach to heavy 
quark production, the perturbative production at next-to-leading order contains terms depending on $\log(p_T/m_Q)$, which need to be 
resummed when $p_T \gg m_Q$. Working with ``massless" fragmentation into hadrons can give the correct results at high $p_T$ while now 
neglecting terms proportional to $\log (m_Q/p_T)$. Cacciari et al. effectively merged the results at fixed order in perturbation theory (accurate at low $p_T$) and next-to-leading log order (accurate at high $p_T$) \cite{Cacciari:1998it}. 

We use \textsc{pythia8.1} to sample the six-dimensional momentum distributions of $Q\bar{Q}$ systems. This works only at leading-order in 
perturbation theory, as almost all state-of-the-art event generators do not include higher orders; including all of the elements in \textsc{pythia} at higher orders is problematic because of increases in the complexity of phase spaces. However, some part of the higher-order behavior is considered by setting the total cross section of charm by hand (we use 
0.6 $\mu b$ for proton-proton collisions at RHIC and 3.0 $\mu b$ for the LHC). The nuclear shadowing is taken into account using \textsc{eks98} nuclear parton distribution functions, and isospin effects are taken into account by sampling from p+p, n+p, p+n, and n+n collisions.

Finally, the spatial dimensions of the phase-space distribution for the $Q\bar{Q}$ states must be sampled. We use the Glauber model to determine the initial centers of mass of the pairs in the transverse plane. An important additional consideration is required by the non-zero 
thermalization time $\tau_0$ in heavy-ion collisions.  We evolve the heavy quarks during this time according to the equations of motion determined by the zero-temperature Cornell potential:

\begin{equation}
\frac{dx^i}{dt}=p^i/E {\rm ,} \; \; \; \;
\frac{dp^i}{dt}=-\frac{\partial V_{{\rm Cornell} }}{\partial x_i}{\rm .}
\end{equation}
These equations are not {\it manifestly} Lorentz-invariant due to the frame-dependent $V_{{\rm Cornell}}$, however, it is defined in the 
center-of-mass frame of the pair for reasons in Section \ref{hadronization}, making the equations of motion Lorentz-invariant while neglecting
radiative effects. The equations of motion in the following section generalize these for temperature-dependent potentials and diffusive effects.
The leapfrog method is used to achieve sufficient numerical accuracy: simpler discretizations of the equations of motion were found 
to have unacceptably large rounding errors at forward rapidities.

\subsection{Heavy quarks at finite temperature: Langevin dynamics}
\label{evolution}

Radiative ``splittings" of high-energy partons, which are a significant source of in-medium evolution for high-$p_T$ massless partons, are 
suppressed for heavy quarks because of the ``dead-cone" effect, already known from jet physics \cite{Dokshitzer:1991fd}. The evolution of heavy quarks at finite temperature is approximated as being caused by elastic collisions in the hard thermal loop (HTL) approximation of QCD.
At the temperatures accessible in heavy-ion collisions, thermal heavy quarks exist in the ``diffusive limit" of dynamics. To 
understand qualitatively why this is so, consider a heavy quark with mass $M$, with $M \gg T$. In the hard thermal limit of QCD, the heavy quark suffers kicks leading to momentum transfers on the order of $gT$. The typical momentum transfer is therefore much smaller than the heavy quark mass, meaning that it takes many kicks to change significantly the momentum of the heavy quarks. The dynamics of heavy quarks undergoing 
elastic processes can thus be approximated as diffusive, and therefore {\it differential} in time.

Many general properties can be calculated directly from the Langevin equation. In the rest frame of the medium,
\begin{eqnarray}
\label{langevin}
\frac{dp_i}{dt} &=& -\eta(p_i) p_i +\xi_i(\vec{p}), \nonumber \\
 \left \langle \xi_i(t) \xi_j (0) \right \rangle &=& \left[ (\delta_{ij}-\hat{p}_i \hat{p}_j) \kappa_T(p) 
 + \hat{p}_i \hat{p}_j  \kappa_L(p) \right] {\rm ,} \nonumber \\
\end{eqnarray}
where $\eta$ is a drag coefficient, $\xi_i$ the stochastic force on the heavy quark, and $\kappa_L$ ($\kappa_T$) the longitudinal (transverse) momentum diffusion coefficient. Often we will approximate $\kappa = \kappa_T = \kappa_L$, which is true non-relativistically. 
These calculations, along with the requirement that an ensemble described by this 
equation thermalizes, leads to the Einstein relation $\eta = \kappa/2MT$. However to simulate individual events, being able to calculate 
ensemble averages only is inadequate; this stochastic differential equation needs to be discretized. The central limit theorem provides the 
means to do this: over a sufficiently long timestep $\delta t$, the {\it sum} of many random kicks is described by a Gaussian with 
width $\sqrt{\kappa \delta t}$. In the end, we use Monte Carlo also in solving the in-medium evolution equation.

In the diffusive approximation, one physical parameter needs to be determined: $D_{HQ}$, the spatial diffusion coefficient for the heavy 
quark. It can be related to $\kappa$ and $\eta$ by examining the expectation value of $\left \langle (x(t)-x(0))^2 \right \rangle = 6Dt$ with the 
Eq. (\ref{langevin}), and can be calculated in the non-relativistic limit as caused by a t-channel gluon exchange in the hard thermal approximation of QCD \cite{Moore:2004tg}, which has been done at next-to-leading order \cite{CaronHuot:2007gq}. It has also been calculated in supersymmetric large-$N_c$ QCD at strong coupling using gauge-gravity duality \cite{CasalderreySolana:2006rq}, as well as in gravity models of the dual for QCD \cite{Mia:2009wj}. The gravity calculations, as well as the poor convergence of HTL effective theory, point to a small diffusion coefficient, $\sim 1/ 2\pi T$. In the simulations that follow we set $2\pi T D_{HQ} = 3.0$ and do not consider here the effects of tuning this parameter, which has large effects on open heavy flavor but smaller effects on the total quarkonium yields.

The dead-cone effect causes brehmsstrahlung to be suppressed in the cone within an angle $~p_T/E$ around the direction of a heavy 
quark's momentum. We would expect this cone to shrink at high momenta, causing the dead-cone effect not to be significant and the effect 
of brehmsstrahlung to be no longer negligible, and indeed, the momenta accessible at the LHC are well into this range. For this reason, a 
future paper will determine the effect of brehmsstrahlung on heavy quarks, as well as describe the heavy quarks at lower momenta without 
the diffusive approximation.


There is one final consideration for the dynamics of heavy quarks that we need to make: the self-interaction of $\bar{Q}Q$ color singlets at 
finite temperature is not negligible, as can be seen with reliable lattice calculations of the correlation between two Wilson lines at finite 
temperature \cite{Kaczmarek:2003ph}. The effect of temperature on the potential cannot be neglected; in fact, the earliest prediction of significant $J/\psi$ suppression was based entirely on changes in this potential with temperature.

This raises another physical question: what thermodynamic potential of $\bar{Q}Q$ color singlets is appropriate to use, the internal energy used 
for adiabatic processes, or the free energy used for isothermal processes? This question has been investigated by comparing 
potential model results for quarkonium correlation functions with the results from lattice QCD 
\cite{Mocsy:2007bk, Petreczky:2010tk}. It remains difficult to distinguish which potential is best for quarkonium at finite temperature. We continue to use the internal energy as a function of temperature, as was done in \cite{Young:2008he, Young:2009tj}, based on the reasoning in 
\cite{Shuryak:2004tx} comparing the timescales of binary bound states, related to the binding energies of these states, to the relaxation times of the medium at finite temperature.

\subsection{Heavy quark hadronization}
\label{hadronization}

Heavy quarks hadronize into mesons with either one or two heavy quarks. Because heavy quarks are produced almost entirely in 
flavor-conserving processes, heavy quarks are produced with an anti-quark of the same flavor and both of these possibilities for hadronic 
final states need to be considered, particularly at center-of-mass energies near thresholds for quarkonium; the production of a given hadronic 
state should never be considered singly.

First, consider open heavy 
flavor: a single heavy quark $Q$ with momentum $P$ fragments into a hadron $Q\bar{q}$ with momentum $zP$, leading to an energy 
difference $\Delta E$ between the initial and final state. General quantum-mechanical arguments lead one to estimate the probability for 
a given transition to be $\propto 1/(\Delta E)^2$, leading to the Peterson fragmentation formula \cite{Peterson:1982ak}:
\begin{equation}
f(z) \propto 1/[z(1-1/z-\epsilon_{ {\rm HQ}}/(1-z))^2] {\rm .}
\end{equation}
In \textsc{martini}, whenever heavy quarks are determined to lead to open heavy flavor mesons, this function is sampled to determine the 
momentum fraction of the resulting meson. At this point in time, for simplicity, the various open heavy flavor states are not distinguished.


The previous argument assumed that the heavy quark was sufficiently energetic (and slightly off-shell) so that a final open flavor state was 
kinematically accessible. Considering the possibility that this state is not accessible allows one to estimate the production of hidden 
heavy flavor states.  Given a heavy quark and anti-quark, with momenta $p_1$ and $p_2$, respectively, whose spatial separation in the lab frame is given by $r_{12}$, we define the modified invariant mass of the pair to be
\begin{equation}
M_{12}=\sqrt{-(p_1+p_2)^2}+V(r_{12}) {\rm ,}
\label{invariantMass}
\end{equation}
where $V(r)$ may be reasonably approximated at chemical freezeout with the Cornell potential. This invariant mass determines what final states 
are kinematically accessible to the pair if it hadronizes adiabatically; if $M_{12}<2m_{D^+}$, open heavy flavor would be kinematically inaccessible to the pair and the final state must be quarkonium. 

In this way, our method of hadronization resembles the color evaporation model \cite{Fritzsch:1978kn, Amundson:1996qr}, which examines simply the invariant mass of the pair to determine whether or not quarkonium is formed. The color evaporation model is quite successful in determining the quarkonium yields in experiments using a range of targets and projectiles, however, it makes no predictions concerning the important observable of $J/\psi$ polarization. The color singlet model \cite{Baier:1981uk} and applications of NRQCD \cite{Cooper:2004qe, Chung:2009xr} make accurate predictions of the $J/\psi$ yields only with some tuning but after being calculated to higher orders in perturbation theory are yielding better agreement while describing polarization.

For the purposes of this paper, the distinctions between the mesons are often ignored. Specifically, Peterson fragmentation leads to the 
production of $D$ mesons, without $D^+$, $D^-$, and $D^*$ mesons distinguished, and while the color evaporation model can use the invariant masses of the pairs to separate the quarkonium into various excited states, there is no distinction made between $\eta_c$ and the three polarizations of $J/\psi$. These states can have similar masses but different quantum numbers leading to different decay channels. For this reason, the final yields calculated with \textsc{martini} need to be multiplied by well-defined correction factors before being compared with data, for example the yields of ground-state quarkonium, when being compared with $J/\psi$ yields, needs to be multiplied by a factor of 3/4 so that the $\eta_c$ particles are not included in the yields. For some processes, for example $e^+e^-$ annihilation, such a factor would be inappropriate 
(the $\eta_c$ has the wrong quantum numbers for this process and is not produced at all), however in heavy-ion collisions these quantum 
numbers are best estimated as entirely randomized, leading to each state's yields being proportional to its degeneracy.

Heavy-ion collisions lead to the possibility of {\it recombinant} quarkonium production. Because of the masses of heavy quarks, there is at leading order only one hard process leading to heavy quark production per proton-proton collision; the production of multiple heavy quarks is subleading in $\alpha_s$. However in a heavy-ion collision, there can be many such hard processes: in a typical central gold-gold collision at RHIC $\sim$20 heavy quark pairs are produced and at the LHC $\sim$100 pairs. This forces us to consider the possibility that a heavy quark from one hard process and an anti-quark from a different hard process might have an invariant mass below the threshold of $2m_{D^+}$ and may form quarkonium.

To determine recombinant production, we do not assume that the heavy quarks hadronize adiabatically; instead the bulk of the collision is 
assumed to act like a heat bath at temperature $T_{ch}$, the chemical freezeout temperature for quarkonium. The possible final states for the 
heavy quarks are assumed to be open heavy flavor mesons and quarkonium (baryons with open heavy flavor, while interesting, are subleading 
and not considered here). Each possible ``pairing" of the $2N$ heavy quarks and anti-quarks is determined by some $\sigma$, an element 
of the permutation group $S_N$. Finally, we associate with each of these pairings an energy:
\begin{equation}
E_{\sigma} = \sum_i^N M_{i \; \sigma(i)}{\rm ,}
\end{equation}
where $M_{i \; \sigma(i)}$ is determined as in Eq. (\ref{invariantMass}). Now that both an energy for each of these pairings and a temperature are 
defined, we can simply weight each pairing by its Boltzmann factor to determine the yields for recombinant production.

The previous argument applies well for an infinite volume cooling uniformly, however in heavy-ion collisions, this limit is not appropriate for 
estimating recombinant quarkonium production. Heavy quarks reach their chemical freezeout surface at different times, and it is appropriate for 
a heavy quark $Q$ to hadronize based on the phase space of heavy quarks {\it at the time Q hadronizes}. For this reason, the hadronization of 
heavy quarks is determined in \textsc{martini} according to this three step algorithm: 
\begin{itemize}
\renewcommand{\labelitemi}{$\bullet$}
\item When $T<T_{ch}$ for a heavy quark $Q$, the modified color evaporation model is used to determine whether or not it forms quarkonium 
with the other heavy quark created in the perturbative process that created $Q$. If quarkonium is formed, then the hadronization stops here.
This channel for quarkonium production is possible for heavy quark pairs produced in regions of the transverse plane at temperatures both 
above and below the deconfinement phase transition, and takes into account the heavy quark pairs that are not significantly 
affected by in-medium dynamics.
\item If quarkonium has not formed, then recombinant production is considered. All oppositely charged heavy quarks are checked with the 
color evaporation model for recombinant production with $Q$ and all such heavy quarks are added to the list. If the set is non-empty, then 
one of these heavy quarks is sampled and a quarkonium state is formed with $Q$. In this situation the Boltzmann factor for open heavy 
flavor production is so small that we neglect it. This channel for production is only available to heavy quarks that exist in QGP. This channel is 
also not available for the directly produced quarkonium (quarkonium formed in the previous hadronization step) as it is approximated as 
kinematically unavailable to the heavy quarks in these states.
\item If neither surviving or recombinant quarkonium has formed, the Peterson fragmentation function for $Q$ is sampled and an open heavy 
flavor meson is formed.
\end{itemize}


Notice that we use two thermodynamic limits for heavy quark hadronization: adiabatic for Peterson fragmentation and color evaporation for 
direct production, and isothermal for recombinant production. While the use of these different limits is well-motivated physically, more work in 
this area can and should be done, especially since all hadronic observables are sensitive to this final step.

\section{Results for RHIC and the LHC}
\label{observables}

We now apply \textsc{martini} to observables at RHIC and the LHC. To maximize the understanding of these effects, we will not consider feed-down of excited quarkonium or $B_c$ mesons into ground states of quarkonium. Vertex detection will allow semi-leptonic weak decays of $B_c$ and other b-hadrons to be observed relatively easily, and these hadrons to be identified, while the use of tracking should allow di-particle invariant mass distributions besides the dielectron invariant mass spectrum to be used to find excited quarkonium; for example, the $\gamma -J/\psi$ invariant mass spectrum can be used to determine the yields of $\chi_c$ in heavy-ion collisions. Indeed, one of our main points in this paper is that  {\it this analysis should be undertaken, considering the high statistics for charm yields at the LHC}.

\subsection{Prompt and non-prompt $J/\psi$ suppression}

\begin{figure}
 \includegraphics{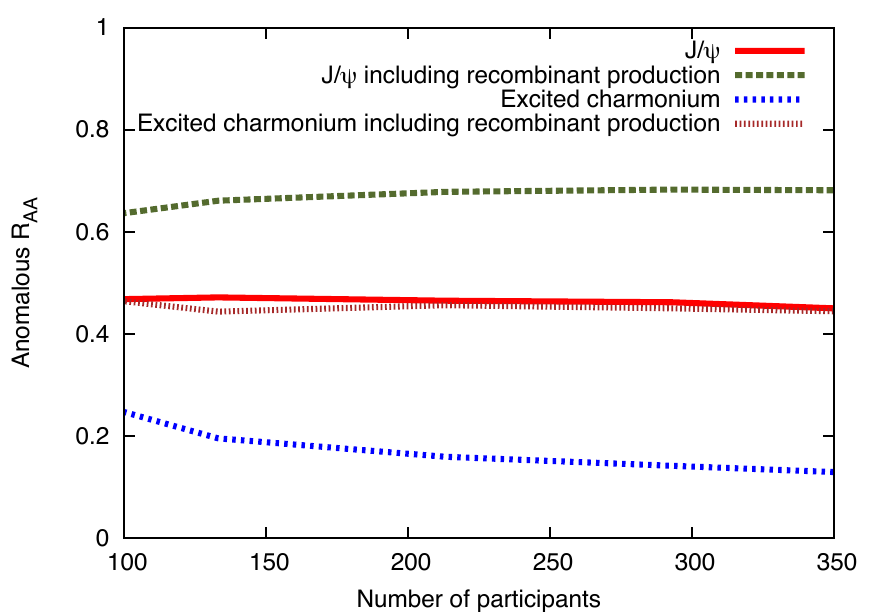}
\caption{The anomalous suppression of charmonium in Au+Au collisions at RHIC energies, with cuts in pseudorapidity appropriate for the 
PHENIX detector ($|\eta|<0.35$). Note the sequential behavior of the suppression, and how event generation leads to larger recombinant production than is predicted by semi-analytic results. 
}
\label{R_AA_quarkonium_RHIC}
\end{figure}

Fig. \ref{R_AA_quarkonium_RHIC} shows the results of \textsc{martini} following from all of the above considerations, applied to gold-gold 
collisions at $\sqrt{s}_{NN}=200 \; {\rm GeV}$. The only cut applied to the final-state particles is that the absolute value of its pseudorapidity 
$|\eta|<0.35$. $R_{AA}$ denotes the ``anomalous" nuclear modification factor: a significant suppression is caused by inelastic scattering of 
quarkonium with the nucleons in the original gold nucleus, and this is not taken into account.

The pattern of suppression in these plots is consistent with sequential quarkonium dissociation: excited quarkonium is destroyed more efficiently than is quarkonium in the ground state, due to the ground state's large binding energy relative to the temperature 
\cite{Karsch:2005nk}. However, this suppression pattern appears here without enforcing the sequential dissociation above predetermined 
dissociation temperatures for the quarkonium states by hand; this pattern occurs simply because less energy has to be transferred from the medium to a $Q\bar{Q}$ system with a small binding energy to excite it above the $2m_D$ threshold than needs to be transferred to a more more tightly-bound $Q\bar{Q}$ system. Also noteworthy is the size of the recombinant contribution to production. It is larger than was estimated in \cite{Young:2009tj}. Simple estimates of recombinant production fail to take into account the strong correlation of charm quarks in position, because of their scaling with number of collisions in the transverse plane which peaks sharply at the center. Also, \textsc{martini}, by simulating the full event, correctly fluctuates the number of charm 
quarks per event and considers events with more charm quarks than average, taking into account the large recombinant production possible in those events.

\begin{figure}
 \includegraphics{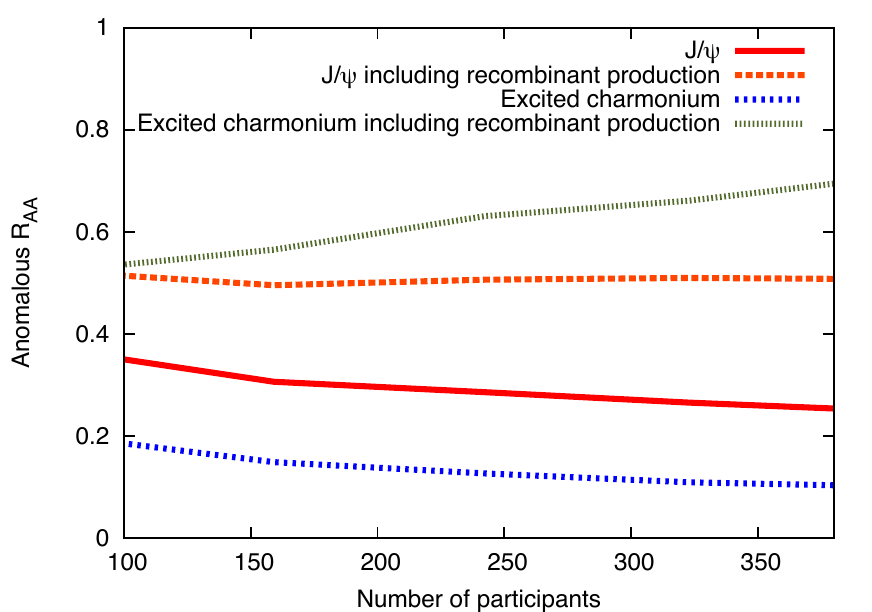}
 \caption{The anomalous suppression of quarkonium for Pb+Pb collisions at LHC energies, with cuts appropriate for the CMS detector. 
 }
\label{R_AA_quarkonium_CMS}
 \end{figure}

Fig. \ref{R_AA_quarkonium_CMS} shows the same algorithms applied to lead-lead collisions at $\sqrt{s}_{NN}=2.76 \; {\rm GeV}$. 
Recombinant production of excited charmonium leads to an increase in $R_{AA}$ with $N_{\rm part}$, due to large charm densities at 
small rapidity. Note especially how including recombinant production of excited states leads to a larger $R_{AA}$ than that of the $J/\psi$ yields. The recombinant production explains why the sequential suppression pattern was not observed by CMS.  We predict that if the nuclear ``absorption" cross-section for charmonium is taken into account properly as in \cite{Adare:2011yf}, the CMS results could use the proper 
baseline for the initial quarkonium production and would show this increase in non-prompt $J/\psi$ $R_{AA}$. Such a characterization of very many cold nuclear matter effects can best be understood with accompanying $pA$ data.

\begin{figure}
 \includegraphics{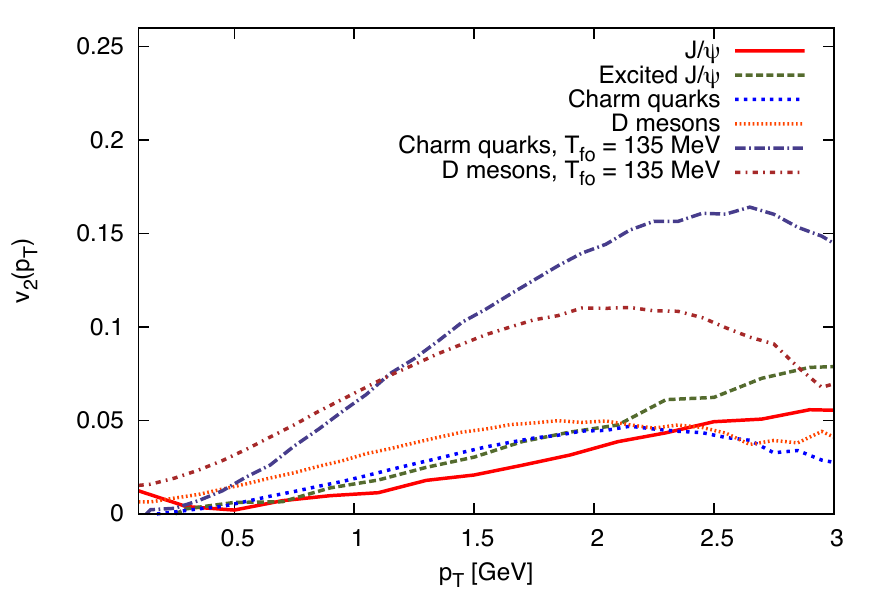}
 \caption{$v_2(p_T)$ for charm mesons in a Au+Au collision with $\sqrt{s_{NN} } =200 \; {\rm GeV}$, when $b=5 \; {\rm fm}$.
 }
 \label{v2_JPsi_PHENIX}
 \end{figure}

\begin{figure}
 \includegraphics{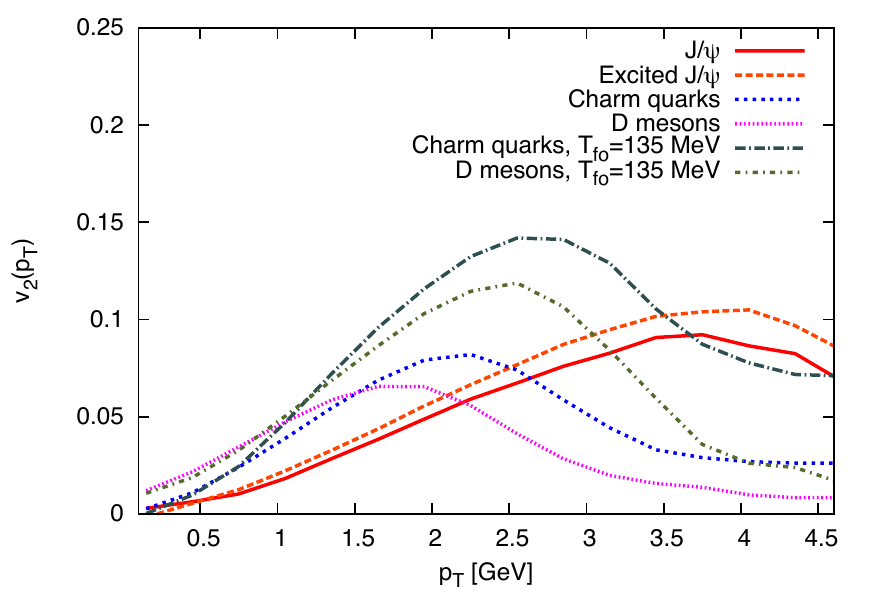}
 \caption{$v_2(p_T)$ for charm mesons in a Pb+Pb collision with $\sqrt{s_{NN}}=2.76 \; {\rm TeV}$, when $b=5 \; {\rm fm}$.
 }
\label{v2_JPsi_CMS}
 \end{figure}

Using \textsc{martini}, heavy quark flow can be re-examined. Fig. \ref{v2_JPsi_PHENIX} shows $v_2(p_T)$ for charm both before and 
after hadronization, and for two kinetic freezeout temperatures, $T_{\rm fo}=190\; {\rm MeV}$ and $T_{\rm fo}=135\; {\rm MeV}$. The $v_2$ result for open charm with the high kinetic freeze-out temperature is smaller than the result in \cite{Moore:2004tg} for 
$2\pi T D_{{\rm HQ}} = 3$ by about a factor of one half, for a simple reason: the flow of heavy quarks develops slower than the flow of the bulk, 
so that quicker freeze-out leads to smaller azimuthal anisotropies. The smaller kinetic freeze-out temperature for open charm leads to larger 
azimuthal anisotropy, and therefore, sequential kinetic freeze-out can explain the different results for the flow of open and hidden charm. A simple argument related to the binding energy of mesons, as well as the results from gauge-gravity duality \cite{Dusling:2008tg}, suggest that the kinetic freezeout temperature for quarkonium should be larger than the freezeout temperature for open heavy quark mesons. In this way, our plot shows how a small $v_2(p_T)$ for quarkonium, relative to the $v_2$ for open heavy flavor, can be explained. Fig. \ref{v2_JPsi_CMS} shows our results for heavy quarkonium flow at the LHC, compared with the flow 
of single charm mesons at the same kinetic freezeout temperature. Finally, note the important effect of fragmentation on the $v_2(p_T)$ results 
for open charm mesons.

\subsection{$B_c$ mesons}

Schroedter et al. predicted $B_c$ meson production in heavy-ion collisions, estimating the production to be one such meson every 20 
collisions \cite{Schroedter:2000ek}. Detecting these states would be difficult with the results from the original incarnations of the detectors at RHIC: the vertices of their decays were not detected, instead the momenta of the relatively stable results of their weak and electromagnetic decays are measured, analyzed, and binned according to pair invariant masses to determine their yields. This makes the progenitors of two-body decays (such as the $J/\psi$) difficult enough to observe, and makes the observation of $B_c$ mesons by their three-body decay
challenging.

This situation should change with vertex detection at RHIC as well as at the LHC. The capability of CMS in this regard had already been 
demonstrated in distinguishing $\chi$ from $J/\psi$ mesons in proton-proton collisions. To date, the results for quarkonium yields in heavy-ion 
collisions were made using the dielectron spectra from these collisions; this leads to all results suffering from the feeddown caused by excited 
states and makes the determination of the yields of particles which do not have simple decays into dileptons challenging at best.

\begin{figure}
 \includegraphics{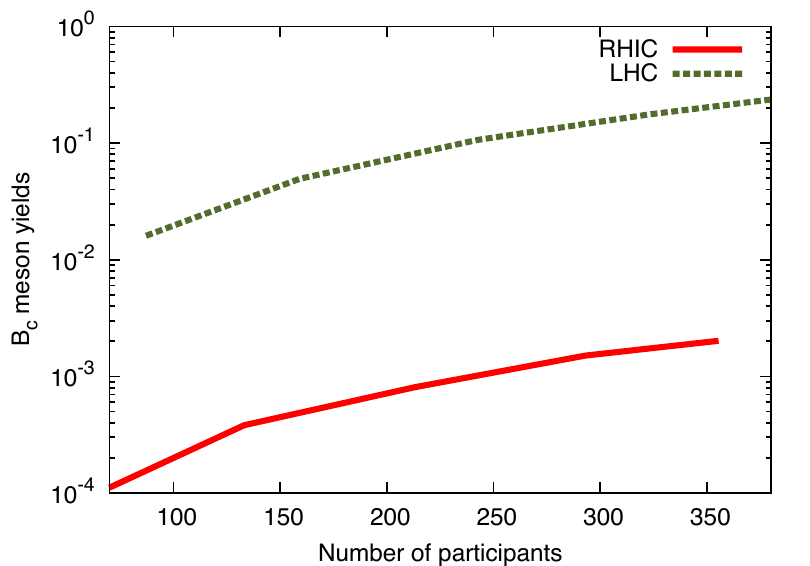}
\caption{ Predicted $B_c$ meson yields at both RHIC and LHC energies, in the pseudorapidity ranges of the PHENIX and CMS detectors, 
respectively, as a function of centrality.
}
\label{BcYields}
 \end{figure}

Fig. \ref{BcYields} shows \textsc{martini}'s prediction for $B_c$ meson yields, at both RHIC and LHC energies, for the pseudorapidity ranges 
detectable by the PHENIX and CMS detectors, respectively. It is worth noting that according to these predictions, heavy-ion experiments can be the main means for studying these rare states. Also, as these states are produced entirely recombinantly, they are a unique probe of finite-temperature hadronization.

\section{Conclusions}

Using full event generation of heavy-ion events, we have examined and re-examined heavy quark observables at RHIC and the LHC. We 
have worked completely in the diffusive limit of the heavy quark dynamics: the quarks were described with Langevin equations of motion 
with a potential term, derived from lattice QCD simulations. Hadronization involved an elaboration of the usual color evaporation model. 
Full event generation allowed for reliable estimation of recombinant quarkonium production, including the production of $B_c$ mesons. We 
found that this recombinant production is larger than previous estimates.

The first improvement that needs to be made here is the elimination of the ``diffusive approximation" for heavy quarks. In order to do so, we can calculate the differential HTL rates $d\Gamma(p)/d^3 q$ and sample these quite easily. However, the radiative rates are also important for heavy quarks at the LHC, where transverse momenta are accessible for which the dead-cone effect is less significant. Work in progress is eliminating the diffusive approximation from these simulations.

\section{Acknowledgments}

CY especially thanks J. Matthew Durham, for providing important input and explanations about detector capabilities. CG, SJ, and CY were supported by the Natural Sciences and Engineering Research Council of Canada and BPS was supported in part by the US Department of Energy under DOE Contract No. DEAC02-98CH10886 and by a Lab Directed Research and Development Grant from Brookhaven Science Associates.

\end{document}